\documentclass[11pt]{article}

\usepackage{amsfonts,amsmath,amssymb}
\usepackage{epsfig}
\usepackage{boxedminipage}

\newcommand{\ket}[1]{| #1 \rangle}

\renewcommand{\phi}{\varphi}

\newcommand{\ceil}[1]{\left\lceil #1 \right\rceil}

\renewcommand{\to}{\hspace{2mm}\longrightarrow\hspace{2mm}}

\renewcommand{\mod}[1]{\hspace{0.05mm}\left(\text{mod }#1\right)}

\newcommand{\CNOT}{\textnormal{CNOT}}
\renewcommand{\,}{\hspace{1mm},\hspace{1mm}}
\newcommand{\NOT}{\text{NOT}}
\newcommand{\TOFFOLI}{\text{Toffoli}}
\newcommand{\inc}{\text{inc}}
\newcommand{\cinc}{\text{c-inc}}

\begin{document}
\title{\textbf{Reversible addition circuit using one ancillary bit with application to quantum computing}}
\author{Phillip Kaye \thanks{prkaye@iqc.ca,
School of Computer Science, University of Waterloo, Waterloo, ON,
Canada.} }

\maketitle

\begin{abstract}
Most of the work on implementing arithmetic on a quantum computer
has borrowed from results in classical reversible computing (e.g.
\cite{VBE95}, \cite{BBF02},\cite{DKR+04}).  These quantum networks
are inherently classical, as they can be implemented with only the
Toffoli gate.  Draper \cite{D00} proposed an inherently ``quantum''
network for addition based on the quantum Fourier transform.  His
approach has the advantage that it requires no carry qubits (the
previous approaches required $O(n)$ carry qubits).  The network in
\cite{D00} uses quantum rotation gates, which must either be
implemented with exponential precision, or else be approximated. In
this paper I give a network of $O(n^3)$ Toffoli gates for reversibly
performing in-place addition with only a single ancillary bit,
demonstrating that inherently quantum techniques are not required to
achieve this goal (provided we are willing to sacrifice quadratic
circuit depth).  After posting the original version of this note it
was pointed out to me by C. Zalka that essentially the same
technique for addition was used in \cite{BCD+96}.  The scenario in
that paper was different, but it is clear how the technique they
described generalizes to that in this paper.
\end{abstract}

\section{Introduction}
Quantum algorithms for factoring and discrete logarithms require the
ability to perform arithmetic operations on a quantum computer.  One
such operation is the addition of two $n$-bit numbers.  In this
paper I focus on the following problem.  Suppose we have two $n$-bit
registers containing the binary representations of two $n$-bit
numbers $a$ and $b$.  We wish to compute the $n$ least significant
bits of $a+b$ (i.e. we want to compute $a+b\mod{2^n}$). We want the
result to be computed in-place of the register initially containing
$b$.  Specifically, we are interested in a circuit implementing
\[a\,b\to a\,a+b\mod{2^n}.\]
Furthermore, we want the the circuit to be \emph{reversible}.  A
reversible circuit for the above task could be directly used in a
quantum network to perform
\[\ket{a}\ket{b}\to \ket{a}\ket{a+b\mod{2^n}}.\]
Several quantum circuits have been proposed that perform this task.
Most of these are inherently classical reversible circuits
(\cite{VBE95}, \cite{BBF02},\cite{DKR+04}) and they require at least
$n$ ancillary bits to keep track of the carry.  For most of the
practical proposed schemes for implementing quantum computers,
qubits will be a very ``expensive'' resource.  Thus there is
significant interest in implementing operations using as few qubits
as possible.  Draper [D00] has described a quantum circuit based on
the quantum Fourier transform that performs addition using no
ancillary qubits. This approach uses rotation gates that must be
implemented with exponential precision, or else be approximated. It
is therefore of interest to investigate whether addition can be done
without the need for $O(n)$ ancillary bits, in a classical
reversible way that does not require these quantum rotation gates.
In the following sections, I will describe such a reversible circuit
for addition that requires only 1 ancillary bit, and $O(n^3)$
Toffoli gates.   After posting the original version of this note it
was pointed out to me by C. Zalka that essentially the same
technique for addition was used in \cite{BCD+96}.  The scenario in
that paper was different, but it is clear how the technique they
described generalizes to that in this paper.

\section{Controlled NOT gates}
\begin{figure}[h]
\begin{center}
\epsfig{file=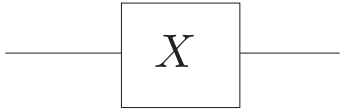, height=1.5cm}\caption{\small{The NOT
gate.}}\label{fig_not}
\end{center}
\end{figure}

The addition circuit will be described in terms of NOT gates that
are controlled on various patterns of control bits. The NOT gate,
shown in Figure \ref{fig_not}, simply flips the binary value of a
bit. That is, if the input to the NOT gate is 0, the output is 1,
and vice versa. The quantum version of the NOT gate is represented
by the Pauli $X$ operator, and acts on qubit states in quantum
superposition according to
\[a_0\ket{0}+a_1\ket{1}\overset{\text{X}}{\to}a_0\ket{1}+a_1\ket{0}.\]

\begin{figure}[h]
\begin{center}
\epsfig{file=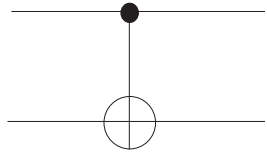, height=2cm}\caption{\small{The CNOT gate.}}\label{fig_cnot}
\end{center}
\end{figure}

The simplest controlled NOT gate is the well known CNOT gate,
depicted in Figure \ref{fig_cnot}.  The CNOT gate applies the NOT
operation to the ``target bit'' (the lower bit in the figure),
conditioned on the ``control bit'' (the upper bit in the figure)
being equal to 1.  If the control bit is equal to 0,
the target bit is left alone. The quantum version of the CNOT gate respects
quantum superpositions and performs
\begin{equation*}
\underset{\text{\begin{small}control\end{small}}}{\underbrace{(a_0\ket{0}+a_1\ket{1})}}
\underset{\text{\begin{small}target\end{small}}}{\underbrace{(b_0\ket{0}+b_1\ket{1})}}
\overset{\text{CNOT}}{\to}
a_0b_0\ket{0}\ket{0}+a_0b_1\ket{0}\ket{1}+a_1b_0\ket{1}\ket{1}+a_1b_1\ket{1}\ket{0}.
\end{equation*}

\begin{figure}[h]
\begin{center}
\epsfig{file=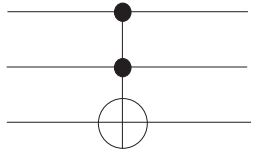, height=2cm}\caption{\small{The Tofolli
gate.}}\label{fig_toffoli}
\end{center}
\end{figure}

A generalized version of the CNOT gate is the Toffoli gate, shown in
Figure \ref{fig_toffoli}.  The Toffoli gate has two control bits,
and one target bit.  The NOT operation is applied to the target
bit conditioned on \emph{both}  the control bits being equal to 1.
The Toffoli gate has the interesting properties
that it is reversible, is and universal in the sense that any
boolean operator can be simulated using it \cite{FT82}. So a quantum
network for addition that uses only Toffoli gates is actually a
classical reversible circuit.

\begin{figure}[h]
\begin{center}
\epsfig{file=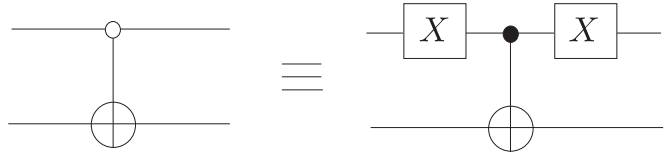, height=2cm}\caption{\small{The
$0$-controlled NOT gate.}}\label{fig_zero_cnot}
\end{center}
\end{figure}

A variant of the CNOT gate is shown in Figure \ref{fig_zero_cnot}.
Here the NOT gate is applied to the target bit conditioned on the
control bit equalling 0. This is indicated in the figure by a hollow
circle on the control bit, as opposed to a solid circle used in
Figure \ref{fig_cnot}. As shown in the figure, the zero-controlled
NOT gate can be implemented using a CNOT, with NOT gates applied to
the control bit before and after the CNOT.

\begin{figure}[h]
\begin{center}
\epsfig{file=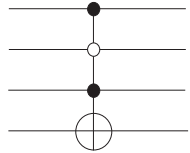, height=2cm}\caption{\small{An example of
a controlled NOT operation which applies the NOT gate to the target
bit, conditioned on three control bits being in the pattern
101.}}\label{fig_fancy_cnot}
\end{center}
\end{figure}

We can also construct more elaborate controlled NOT ``gates'', with
the NOT operation applied conditioned on some $k$ control bits being
in a specified pattern. For example, the ``gate'' shown in Figure
\ref{fig_fancy_cnot} applies the NOT gate to the target bit
conditioned on 3 control bits being in the pattern 101.

\subsection{Complexity of the controlled NOT gates}

I will assume that the ``elementary gates'' we have at our disposal
are $\{\NOT,\CNOT,\TOFFOLI\}$.  To determine the total depth of the
circuits I will consider the circuits to be constructed only out of
these elementary gates.

We have seen how a zero-controlled NOT gate can be simulated using 2
NOT gates and a CNOT gate.  This means that the depth of a
zero-controlled NOT gate is 3.

\begin{figure}[h]
\begin{center}
\epsfig{file=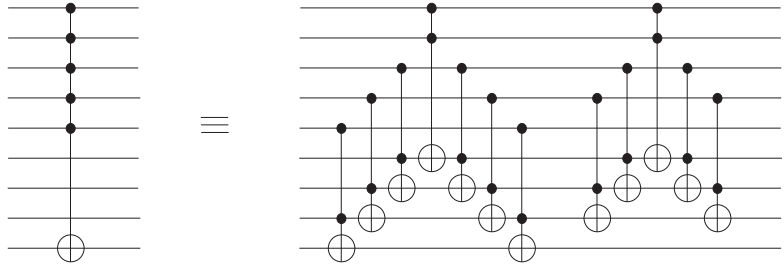, height=3.5cm}\caption{\small{An example of
the construction described in \cite{BBC+95}.}}\label{fig_BBCex}
\end{center}
\end{figure}

Consider $\text{controlled}^k$ NOT gates, similar to that in Figure
\ref{fig_fancy_cnot}, except with all the control bits being
1-controls (solid black dots in the figure).  This gate applies the
NOT operation to the target bit only if all the control bits are in
state $1$.  We are shown how to simulate the $\text{controlled}^k$
NOT gate by Toffoli gates in \cite{BBC+95}.  For a circuit having a
total of $N$ bits, if $N\geq 5$, and  $\ceil{\frac{N}{2}}\geq k\geq
3$, the $\text{controlled}^k$ NOT operation can be simulated using
$4k-8$ Toffoli gates.  Note that the construction in \cite{BBC+95}
does not require dedicated ancillary bits.  The construction is
illustrated for a specific example in Figure \ref{fig_BBCex}.

Consider the requirement that $k\leq \ceil{\frac{N}{2}}$ (i.e. no
more than half of the bits on the circuit can be used as control
bits). The addition circuit I will describe later (Figure
\ref{fig_add}) has $N=2n+1$ bits (where $n$ is the size of each of
$a$ and $b$) and the controlled NOT gates will only use up to $n$
control bits.  So this technical requirement will be satisfied, and
we can use the above result to count the depth of the
$\text{controlled}^k$ NOT gates.

We wish to simulate a more general $\text{controlled}^k$ NOT gates
such as those in Figure \ref{fig_fancy_cnot}, where some of the
control bits are 0-controls (hollow circles).  To do this we have to
add 2 to the overall depth, since we need to conjugate the 0-control
bits with NOT gates, as in Figure \ref{fig_zero_cnot}.  This means
that the depth of a general controlled NOT gate is $4k-6$, where $k$
is the number of control bits (providing the technical requirement
$k\leq\ceil{N}{2}$ is satisfied).

\section{A circuit for incrementing}
\begin{figure}[h]
\begin{center}
\epsfig{file=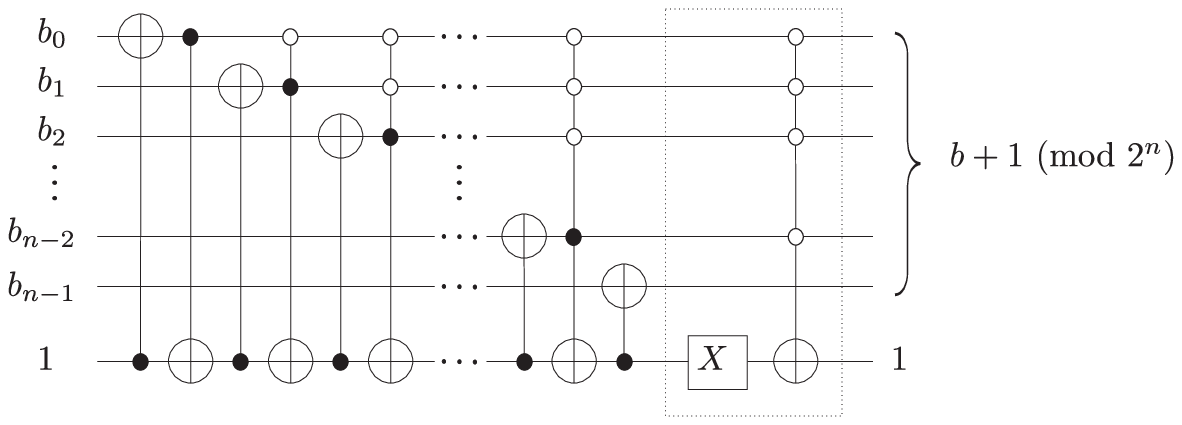, height=5cm}\caption{\small{A circuit
implementing $b\,1\to b+1\mod{2^n}\,1$.}}\label{fig_inc}
\end{center}
\end{figure}

A building block for the addition circuit is a circuit for
\emph{incrementing} an $n$-bit number:
\[b\to b+1\mod{2^n}.\]
The circuit given in Figure \ref{fig_inc} does this with one
ancillary bit, initially set to 1.  The ancillary bit will be known to
equal 1 at the output of the circuit, and so can
be re-used (e.g. for further incrementing).  That is, the circuit in
Figure \ref{fig_inc} implements
\[b\,1\to b+1\mod{2^n}\,1.\]

To understand how the circuit in Figure \ref{fig_inc} performs the
incrementing, imagine how you would increment a binary number ``by
hand''. You would flip the least significant bit.  If you had
flipped it from a 0 to a 1, you would be done.  If you had flipped
it from a 1 to a 0, you would then proceed to flip the next most
significant bit. You would proceed in this fashion until you had
flipped a bit from 0 to 1, at which point you would be done.  The
ancillary bit in the circuit can be viewed as a ``flag'' which
signals the first time you have flipped a bit from $0$ to
$1$, and should stop flipping bits. The flag is $1$ as
long as you should continue flipping bits, and then is set to
$0$ when the condition is reached such that you should stop
flipping bits. We know that at some point we will have flipped the
state of the flag bit from $1$ to $0$, unless the
state of the $n-1$ least significant bits
$(b_{n-2}b_{n-3}\cdots b_0)$ was initially $(11\cdots 1)$.
In this case the state of these bits after incrementing is
$(b_{n-2}b_{n-3}\cdots b_0)=(00\cdots 0)$. So after
incrementing $b$, the flag bit should be re-set to $1$ by a
NOT gate, unless $(b_{n-2}b_{n-3}\cdots b_0)=(00\cdots 0)$.
The portion of the circuit of Figure \ref{fig_inc} surrounded by the
dashed box accomplishes this uncomputing of the ancillary bit.
Note that I originally proposed this incrementing circuit in \cite{KZ04}.

\section{The addition circuit}

\begin{figure}[h]
\begin{center}
\epsfig{file=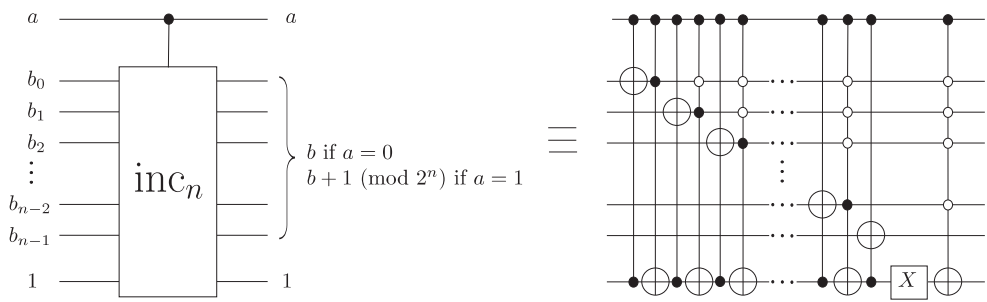, width=13cm}\caption{\small{The controlled
incrementing circuit.}}\label{fig_cinc}
\end{center}
\end{figure}

The addition circuit will make use of a sequence of
\emph{controlled}-incrementing circuits, $\inc_n$.  This will apply
the incrementing circuit to an $n$-bit register containing $b$,
conditioned on a single control bit $a$ equalling $1$. Such a
circuit can be realized by simply adding a control on $a$ to every
gate in the incrementing circuit described above.  The circuit and
its abstract schematic symbol are shown in Figure \ref{fig_cinc}.

\begin{figure}[h]
\begin{center}
\epsfig{file=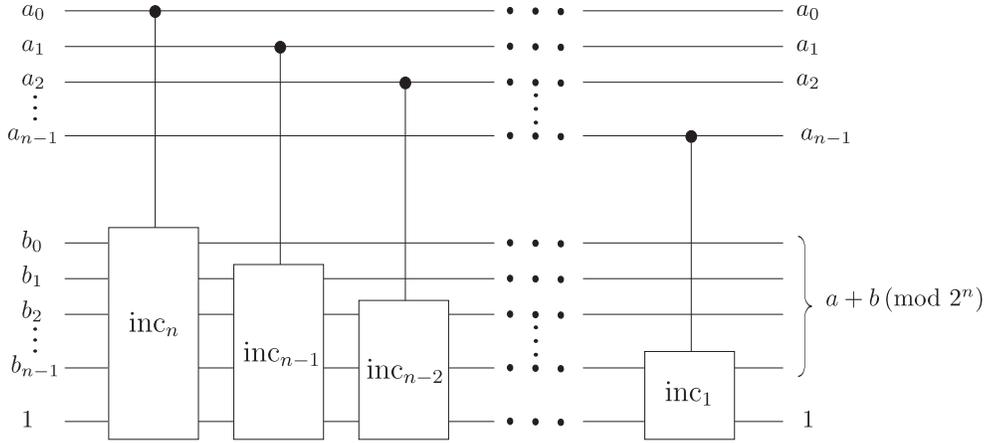, width=13cm}\caption{\small{The addition
circuit.}}\label{fig_add}
\end{center}
\end{figure}

Now it can be seen that the circuit shown in Figure \ref{fig_add}
performs the addition.  The idea is that adding $a$ to $b$ modulo
$2^n$ can be achieved by conditionally incrementing (mod $2^n$) the
$i+1$ most significant bits of $b$, controlled on
$a_i=1$, for $i=n-1,\ldots,0$. The single ancillary
bit is left in the state $1$ at the end of the circuit, and
so can be re-used.

\section{Depth of the addition circuit}

It remains to determine the depth of the addition circuit. The
addition circuit uses a sequence of $n$ controlled incrementing
operations.  Let $|\cinc_k|$ denote the depth of a
controlled-$\inc_k$ operation.  By carefully counting the depth of
each of the general controlled NOT operations in the
controlled-$\inc_k$ circuit, we find
\[|\cinc_k|=\begin{cases}1&\text{ if }k=1\\10&\text{ if
}k=2\\2k^2+k-5&\text{ if }k\geq 3.\end{cases}\] Note that in the
above calculation, I assumed that the controlled-$\inc_k$ operations
will be in the context of a circuit having at least $2k$ bits in
total, so that the technical requirement of \cite{BBC+95} is
satisfied.  This is true in the context of the addition circuit,
which has $2n+1$ bits (where $n$ is the size of $a$ and $b$).

For $n\geq 3$, the overall depth of the addition circuit is
\[\sum_{k=1}^n |\cinc_k|=\frac{2}{3}n^3+\frac{3}{2}n^2-\frac{25}{6}n+8.\]

\section{Conclusion}
We have seen how to perform in-place addition of two $n$-bit numbers
with only 1 ancillary bit by an $O(n^3)$-depth reversible circuit
comprised of NOT, CNOT and Toffoli gates.  After posting the
original version of this note it was pointed out to me by C. Zalka
that essentially the same technique for addition was used in
\cite{BCD+96}.  The scenario in that paper was different, but it is
clear how the technique they described generalizes to that in this
paper.

The circuit in Figure \ref{fig_add} can be directly applied quantum
states, provided we have a quantum implementation of the NOT, CNOT
and Toffoli gates. In the language of quantum computing, the circuit
would perform
\[\ket{a}\ket{b}\to\ket{a}\ket{a+b\mod{2^n}}.\]
The circuit depth of $O(n^3)$ probably means that it would not be practical for
classical (low power) computing \cite{F04}, but it may be practical for quantum computing
if qubits are at a premium.

\end{document}